\documentclass[aps,letterpaper,amssymb,prl]{revtex4}
\usepackage{amssymb}
\usepackage[usenames,dvips]{color}
\usepackage{graphicx}
\usepackage{amsmath}
\usepackage{times}
\usepackage{subfigure}
\usepackage{times}
\begin{document}
\title{Other incarnations of the Gross-Pitaevskii dark soliton}
\author{ Indubala I Satija $^{a,b}$ and Radha Balakrishnan $^{c}$
\footnote {Corresponding author.\\
{\it E-mail address:} radha@imsc.res.in}
}
 \affiliation{$^{a}$ Department of Physics, George Mason University, 
 Fairfax, VA 22030}
\affiliation{ $^{b}$ National Institute of Standards and
Technology, Gaithersburg, MD 20899}
\affiliation{$^{c}$ The Institute of Mathematical Sciences, Chennai 600113, India}
\begin{abstract}

We show that the dark soliton of the
Gross-Pitaevskii equation (GPE) that describes the Bose-Einstein 
condensate (BEC)  density of  a system of {\it weakly}
repulsive bosons, also describes  that  of a system of 
{\it strongly}  repulsive
hard core bosons at {\it half filling}.
As a consequence of this, the  GPE soliton  gets related  to 
the magnetic soliton in an easy-plane ferromagnet,  
where it describes the square of the in-plane magnetization 
of the system.
These relationships are shown to be useful in understanding 
various characteristics of solitons in these 
distinct many-body systems.\\

\end{abstract}
\maketitle

\noindent {\bf 1.  Introduction}\\

The study  of solitary waves and solitons 
has been  one of the most active areas of research 
in nonlinear systems for the past three decades \cite{daux}.
As is well known,  a solitary wave  is a spatially localized
traveling wave which 
 maintains its shape, size and speed as it travels. If a solitary wave
retains all its features even after a collision with another 
solitary wave, it is called a soliton.
Some of the iconic examples of completely integrable  
nonlinear partial differential equations 
arising in applied mathematics  and exhibiting 
such strict soliton solutions \cite{ablo} include
the Korteweg-deVries equation, the sine Gordon equation, 
the nonlinear Schr\"{o}dinger equation, etc. 
However, in physics literature a solitary wave itself is often referred
to as a soliton due to its particle-like properties and  we  will do so 
in this paper as well.  
Solitons appear in diverse fields \cite{daux} such as hydrodynamics, 
fiber optics, biophysics, high energy physics, magnetism 
 and in various condensed matter systems including a 
Bose-Einstein condensate (BEC). 
In addition to theoretical investigations, they have  also been  
observed experimentally  in various nonlinear systems. 

A BEC is an intrinsically nonlinear system.
The Gross-Pitaevskii equation (GPE) \cite{book} 
has been central to explaining various fundamental aspects 
of the BEC of a system of  weakly 
repulsive  bosons.
One of the hallmarks of this system is the dark soliton that has been 
theoretically predicted \cite{tsuz} as a unidirectional 
traveling wave solution for 
the condensate density.  This soliton, which  dies out
when its speed approaches the speed of sound, has also been observed  
experimentally \cite{burg}. The study of solitary wave propagation
in BEC remains an active frontier with emphasis on non-GPE dynamics \cite{kol}
and many body effects such as quantum fluctuations and depletion that cannot be 
described by GPE \cite{lew,Carr}.

In view of the relatively few examples of  systems exhibiting
solitons, any relationship that is found between
different physical systems is fascinating, since it unveils
the universal mathematical aspects  underlying  these
nonlinear  systems that describe completely different
physical phenomena. 
In this paper, we show that the dark soliton of the
GPE that describes the  
BEC  density of  a system of {\it weakly}
repulsive bosons, also describes  that  of a system of 
{\it strongly}  repulsive
hard core bosons at {\it half filling}.
This in turn leads to a connection between  the  GPE soliton  and  
a  magnetic soliton in an easy-plane ferromagnet.  
These relationships are shown to be useful in understanding 
various characteristics of solitons in these 
distinct many-body systems.  

The complex order parameter (i.e., the condensate wave function) 
describing a  BEC  that satisfies the GPE
can be identified with the bosonic coherent state 
average \cite{lang} of the boson 
annihilation operator in the  continuum version of the Bose-Hubbard model. 
Analogously, in a strongly interacting limit described by a system of hard-core bosons (HCB) 
the appropriate order parameter for this system  would be the spin
coherent state average \cite{radha1} 
as the system maps to a pseudospin model \cite{mats}. 
In the continuum description 
of the dynamics of the extended lattice 
Bose-Hubbard model for hard-core bosons 
with nearest neighbor interactions, this order parameter satisfies 
an evolution equation  that is different from the GPE \cite{prl}.
We will refer to this equation as the HGPE, where the prefix 
"H" has been used to denote its connection to HCB. 

Unlike the GPE which describes 
the dynamics of a BEC with no depletion, the HGPE 
evolution encodes both the normal and  condensate fractions
in the bose system.
In our recent study, we showed that the HGPE supports \cite{prl}  both  a  dark soliton  and 
an antidark\cite{anti} soliton 
(i.e., a bright soliton on a pedestal) for the bosonic density. 
In other words, 
the dark soliton of the 
BEC of weakly repulsive bosons acquires a partner that is  bright,
when the repulsion becomes extremely  strong.
Apart from being bright, the anti-dark soliton is found to be 
of quite a distinct variety compared to the dark soliton that
resembles the GPE soliton.
Away from half-filling, i.e., when there is a 
nonvanishing particle-hole imbalance
in the background, 
the dark soliton  dies out when its
propagation speed approaches the speed of  sound (like the GPE soliton), 
while  its brighter partner is found to persist all the way up to  
the sound velocity \cite{prl}.

In contrast, at half-filling, when the particle-hole 
imbalance is zero, the dark and the antidark solitons 
in the particle density profile
 of the HGPE become mirror images of each other, and they both die 
out at the sound velocity. Interestingly, in the corresponding 
condensate density, the two solitons become degenerate and cease 
to exist as separate entities. In this special case, the condensate
density profile is a dark soliton that dies out as its 
speed approaches the speed of sound \cite{prl}. 

In this paper, we show that for the half-filled case, 
 the localized functional form of the  solitary wave solution for 
the  density of the  hard-core bosonic condensate
 described by the HGPE agrees with that of the dark soliton 
of the GPE, to a very good approximation.  
In other words, if one  focuses on the
condensate fraction of the hard-core bosonic system,
 this strongly repulsive bose gas  supports
a solitary wave that   has  almost the same profile as the GPE soliton that
describes  the BEC of a weakly repulsive   bose gas.

Furthermore, since the HCB gas in the spin 
coherent state formulation of 
the extended Bose-Hubbard model with 
nearest neighbor interactions 
mimics  an anisotropic, easy-plane ferromagnetic chain,
the well known dark soliton of the GPE also relates to the 
magnetic soliton in the  easy-plane ferromagnet, where it
describes the profile of the {\it square of the in-plane magnetization}.

In section  2, we begin with the many-body Bose-Hubbard Hamiltonian 
that describes the low-energy behavior
of  bosonic atoms in an optical lattice, and briefly outline some steps 
that lead to the order parameter equations
in the weakly and strongly interacting limits, respectively. In sections 
3 and 4, we first derive the evolution equation for
the traveling waves in the condensate fraction of the atomic cloud 
obtained from the HGPE,
and  show that to a very good approximation,
this equation agrees with the corresponding equation  
derived from  the GPE. In section 5, we discuss 
the relationship between  the density solitons of the BEC 
 and the magnetic solitons of an easy-plane ferromagnet.\\ 

\noindent {\bf 2.  The Bose-Hubbard model,  the GPE,   and  the HGPE}\\

We begin with the extended lattice Bose-Hubbard model in 
$d$ dimensions, whose Hamiltonian is given by 
\begin{equation}
H=-\sum_{j,a}[t \,b_j^{\dagger} b_{j+a}+ V n_j n_{j+a}]\\+\sum_j U n_{j}  
(n_{j}-1) 
-(\mu-2t)n_{j},
\label {BH}
\end{equation}
where  $b_j^{\dagger}$ and $b_j$ are the creation and  
annihilation operators 
for a  boson at the lattice site $j$,  $n_{j}$ is the number operator, $a$ 
labels nearest-neighbor (nn) separation , 
$t$ is the nn hopping parameter, $U$ is the on-site repulsion strength, 
and $\mu$ is the chemical potential.
To soften the effect of strong onsite repulsion, we add
an attractive nn interaction ($V > 0$).
Such a term may mimic certain characteristics of the long range 
dipole-dipole interaction, that has been considered 
in several recent studies \cite{Baiz}.
The term $2tn_j$ is added so that the terms involving $t$ 
reduce to the  kinetic energy expression in the continuum version  
of the many-body  bosonic Hamiltonian.

Conventionally, one defines the order parameter for a 
many boson system 
to be the thermodynamic expectation
value of the boson field operator. Invoking the concept 
of a broken gauge symmetry  allows this expectation value 
to be nonzero below the BEC
transition temperature.
It has been argued\cite{lang} that this 
order parameter may be chosen to 
be the expectation value of  the boson annihilation operator in the 
bosonic coherent state 
(also known as the harmonic oscillator coherent state or
 Glauber coherent state) representation of the pure quantum state.
In this description, it can be shown that  
quantum fluctuations are absent.

The Heisenberg equation of motion for the
boson annihilation  operator determined from 
(\ref{BH}), after a bosonic coherent state averaging
yields the following GPE equation for the condensate 
order parameter $\Psi_g({\bf r}, t)$ in the continuum description.

\begin{equation}
-(\hbar^2/2m) \nabla^{2} \Psi_g+U |\Psi_g|^2 \Psi_g- \mu \Psi_g = 
i \hbar \partial_t \Psi_g.
\label{gpe}
\end{equation}

The GPE  with $U > 0$  provides  a very useful 
characterization of the  various properties of 
the BEC of {\it weakly repulsive}  bosons  
\cite{book}. 
Here the  condensate density $\rho_g= |\Psi_g|^2$  
 satisfies  a continuity equation,
and the system shows no depletion.

 The HCB limit ($U \rightarrow \infty$) of the 
Bose-Hubbard Hamiltonian  (\ref{BH}) 
has emerged as a useful model to describe various 
characteristics of the BEC of a {\it strongly repulsive} bose system. 
The constraint that two bosons  cannot occupy the same site 
can be incorporated in the formulation by using 
field operators that anticommute at the same site but commute at
different sites, thus satisfying
the same algebra
as that of a spin-$\frac{1}{2}$ system. 
By identifying $b_j$ with the spin flip operator 
$\hat{S}_{j}^{+}$, along with 
$n_{j} = \hat{S}_{j}^{+} \hat{S}_{j}^{-}= \frac{1}{2} - \hat{S}_{j}^{z}$,  
the system can be mapped to
the following  quantum XXZ Hamiltonian in a magnetic field:  

\begin{eqnarray}
H_S=-\sum_{j, a} [t\,\, {\bf {\hat{S}}_j \cdot {\bf {\hat{S}}_{j+a}}} - 
g\,\, \hat{S}_j^z \hat{S}_{j+a}^z]-
\sum_j({\textstyle g} - \mu)\,\, \hat{S}_j^z.
\label{spinH}
\end{eqnarray}

 Here $g = (t-V)d$,  where $d$ is the spatial dimensionality.
As we shall show, $g =(t-V) > 0$ (see below Eq. (\ref{cs})).
Hence the HCB system  maps
to a {\it quantum} spin-$1/2$ system with a 
Heisenberg exchange interaction $t$, 
 an {\it easy-plane exchange anisotropy} $g$ and a
transverse magnetic field $h_z = (g-\mu)$ along the z-direction.

This mapping to spins suggests that a 
natural choice for  the condensate order parameter 
$\Psi_{s}$ of the HCB system
is the average of the spin flip operator in the spin 
coherent state representation
\cite{Radc}, i.e., 
\begin{equation}
\Psi_{s} = < \hat{S}^+ >.
\end{equation}
In this representation, it can be shown that \cite{prl}  
the condensate density, 
 $\rho_s =  |\Psi_s|^2 =  < \hat{S}^{-}> < \hat{S}^{+}> $ 
is related to total particle number density $\rho = < \hat{S}^{-} \hat{S}^{+}>$ 
as,
\begin{equation}
\rho_s = \rho ( 1-\rho)
\label{deqn}
\end{equation}

Setting  $\Psi_s = \sqrt{\rho_s} \exp {i \phi}$ in the 
continuum description,
the condensate order parameter $\Psi_s$ and the particle density 
$\rho$ satisfy the following  equations \cite{prl}
\begin{equation}
i \hbar \dot{\Psi_s}
 =  -\frac{\hbar^2}{2m} (1-2\rho) \nabla^2 \Psi_s
-V_e \Psi_s \nabla^2 \rho
+ 2 g \rho \Psi_s-\mu \Psi_s
\label{sgpe}
\end{equation}
\begin{equation} 
\dot{\rho}  = \frac{\hbar}{2m} \nabla\cdot [\rho(1-\rho)\nabla \phi],
\label{sgpe1}
\end{equation}
with the identification  $t a^{2}  = \hbar^2/m $ and $ V a^2 = V_{e}$. 
Equation (\ref{sgpe}) (which can also be written as 
 coupled equations 
for $\rho$ and $\phi$)  will be referred to as HGPE. 
Using the asymptotic value $\rho \rightarrow \rho^{0}$ in 
Eq. (\ref{sgpe}), the chemical potential $\mu$ is found to 
be  $\mu = 2 g \rho^{0}$.

From Eq. (\ref{deqn}), we note that it is important
to distinguish between the  total bosonic particle density $\rho$ 
and the condensate density $\rho_s$, where $\rho=\rho_s+\rho_d$.
Here, $\rho_d$ describes the depletion, i.e.,  the normal 
component of $\rho$.

Linearizion of HGPE by considering small amplitude 
fluctuations in the asymptotic value of $\Psi_{s}$ 
leads to a Bogoliubov-like spectrum.
For small momenta, the spectrum becomes linear, leading to the
 sound velocity $c_{s}$ for the hard-core system as 
\begin{equation}
c_{s} = (2g \rho_{s}^{0}/m)^{\frac{1}{2}},
\label{cs}
\end{equation}
where $\rho_{s}^{0}$ is the asymptotic value of the
condensate density. Since $c_{s}$ 
must be real, we obtain the condition\\  $g = (t -V) > 0$. 

In the next section, we will derive an equation of motion 
for the {\it condensate density }
$\rho_s$ that arises from HGPE, for the  half-filled case,
when the  asymptotic particle density is half, i.e.,
 when the number 
of particles equals the number of holes in the background.\\

\noindent {\bf 3. HCB Condensate density evolution in the 
half-filled case of HGPE}\\

By seeking unidirectional traveling wave 
solutions (along the x-direction) of the total density 
$\rho$ of the form,
\begin{equation}
\rho (z) = \rho_{0}+ f(z),
\end{equation}
where $z=(x-vt)/a$ and $\rho_{0}$ denotes the constant asymptotic value of the 
 background density, it has been shown \cite{prl} 
that the HGPE (Eq. (\ref{sgpe}))
leads to a nonlinear differential equation for 
$f(z)$  which can be solved analytically to a very good approximation, 
to yield  soliton solutions.

Here we focus on the half-filled case
with $\rho_{0} = 1/2$. For this case, the differential equation 
for $f$ takes the form \cite{prl} 

\begin{equation}
\left(\frac{d{f}}{d\bar{z}}\right)^2=4 {f}^2\left[\gamma^2-{f}^2\right].
\label{feqn}
\end{equation}

Here, $\gamma^2 = 1-\bar{v}^2$ with $\bar{v}=v/c_s$. Further,
$\bar{z}= \zeta z$ and $\zeta = \Lambda/\sqrt{1- 2 \Lambda^2}$,
where the microscopic dimensionless parameter 
$\Lambda = c_s/c_ {0}$  
is the speed of sound $c_s$ in the HCB condensate 
measured in units of the zero-point velocity $c_ {0} = \hbar/ma$.

As discussed in \cite{prl}, Eq. (\ref{feqn}) 
provides the following two soliton solutions 
\begin{equation}
f (\bar z)  = \pm (\gamma/2)  \,\, \rm {sech} ( 2 \gamma \bar z) = \pm (\gamma/2)
 \,\, \rm {sech} (z/ \Gamma_{s}),
\label {fsol}
\end{equation}
where  the soliton width $\Gamma_{s} = (2 \gamma \zeta )^{-1}$.
Since $\gamma$ must be real in Eq. (\ref {fsol}), the 
dimensionless parameter $\bar{v}$ must satisfy  $0 < \bar{v} < 1$.

The two solutions for $f$  in (\ref{fsol}) lead to   
a  doublet  of localized solitons, namely, a dark soliton and 
an antidark soliton,   
for the density  of the bosonic cloud $\rho (\bar z) = (1/2) + f(\bar z)$.
Using this  in the  condensate density expression  given in
Eq. (\ref{deqn}), we get
\begin{equation}
\rho_{s}(\bar z) = \frac{1}{4} -f^{2}(\bar z).
\label{fs}
\end{equation}
Writing $\rho_{s}(\bar {z}) = (1/4) + f_{s}(\bar z)$, 
and comparing it with Eq. (\ref{fs}) shows that the variation  $f_{s}$ around
  the asymptotic condensate density and the variation $f$ around
the total density $\rho$ are related as follows:
 
\begin{equation}
f_{s}(\bar z) =   \rho_{s}(\bar z) - \frac{1}{4} = -f^{2}(\bar z)
\label{fsbar}
\end{equation}

Differentiating both sides of Eq. (\ref{fsbar})  with respect to $\bar z$ 
and using Eq. (\ref{feqn}) in the resulting equation, we obtain the following
differential equations for the normalized condensate density 
$\bar \rho_{s} = \rho_{s}/(1/4) $  and the 
corresponding normalized variation $\bar f_{s} = f_{s}/(1/4) $, 
respectively.

\begin{equation}
(\frac{d \bar{\rho}_s}{d\bar{z}})^2 = ( 1-\bar{\rho}_s)^2 ( \bar{\rho}_s-\bar{v}^2 )
\label{eqnrs}
\end{equation}

\begin{equation}
(\frac{d \bar{f}_s}{d\bar{z}})^2 = \bar{f}_s^2 (\bar{f}_s+\gamma^2 )
\label{eqnfs}
\end{equation}

Unlike the doublet soliton solutions of $f$ given in 
Eq. (\ref{fsol}), $\bar {f}_s$ has the  
unique solution
\begin{equation}
\bar{f}_s =- 4 f^{2} =  -\gamma^2 \,\,  \rm {sech}^2\, 2 \gamma \bar {z} = 
 -\gamma^2 \,\, \rm {sech}^2\,  {z/ \Gamma_{s}}.
\label{gammas}
\end{equation}

\begin{figure}[htbp]
\includegraphics[width =1.0\linewidth,height=1\linewidth]{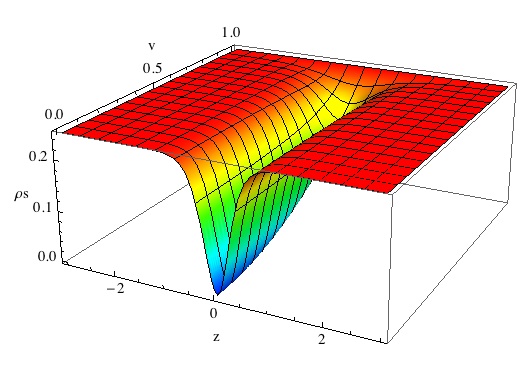}
\leavevmode \caption{(color online) Condensate density  profile
$\rho_{s}$ (Eq. (\ref{fs}))   
of HGPE at half filling with $V/t=1/3$, showing the behavior of the dark soliton 
as its propagation speed $\bar v = v/c_{s}$  increases from $0$ to $1$.
Note that the soliton flattens out at $\bar v = 1$.}
\label{fig1}
\end{figure}

Figure ~1 shows the dark soliton profiles for the 
condensate  density for various values of  $\bar v$. 
As the plot shows, the 
soliton which is dark when $\bar v = 0$, dies out as 
 $\bar v \rightarrow 1$, i.e., 
as its propagation speed approaches the speed of 
sound.\\

\noindent {\bf 4.  Condensate density evolution in the GPE}\\

In this section,we  show that  
Eq. (\ref{eqnrs}), which  describes the  evolution 
of  the condensate 
density of the HGPE for a half-filled system, has the same 
form as the condensate density evolution equation obtained from the GPE.

Following the  earlier studies of solitons in GPE\cite{book}, we write
the condensate wave function $\Psi_g$ in that system as
\begin{equation}
\Psi_g=  \sqrt{\rho_{g}^{0}}\,\,\, ( \psi_r+i\bar{v})
\label{gpeg}
\end{equation}
where $\rho_{g}^0$ is the asymptotic value of the background density.
 $\psi_r$ denotes the real part of $\Psi_{g}$. Its imaginary part
is  $\bar {v} =v/c_{g}$ is the soliton propagation speed measured
in units of the sound speed $c_{g}$ for the GPE. It is given by 
\begin{equation}
c_g = (U \rho_{g}^{0}/m)^\frac{1}{2}.
\label{cg}
\end{equation}
In Eq. (\ref{gpeg}),  $0 < \bar {v} < 1$. 
Further, it can be shown that $\psi_r$ satisfies \cite{book} 

\begin{equation}
\sqrt{2}\frac{d\psi_r}{d w}-(\gamma^2-\psi_r^2) = 0,
\label{real}
\end{equation}

where $\gamma^2 = 1-\bar{v}^2$ and 
$w = (x -vt)/\xi$, where the healing length $\xi$ is given by
the well known expression $\xi =\hbar /\sqrt{2mU \rho_{s}^{0}} 
= \hbar/\sqrt{2}m~c_{g}$. 

In analogy with the HGPE (see below Eq. (\ref{feqn})), 
we define a dimensionless parameter $\Lambda = c_g/c_{0}$, 
which is the sound velocity in the GPE measured in units of the zero-point
velocity. This enables us to write $\xi$ in terms  of $\Lambda$ as
\begin{equation}
\xi =a /\sqrt{2}\, \Lambda.
\label{xi}
\end{equation}
Hence  $w =  \sqrt{2}\, \Lambda (x - vt)/a = \sqrt{2}\, \Lambda z $.
Using this in Eq. (\ref{real})  and 
squaring both sides of this equation, we get
\begin{equation}
 (\frac{d\psi_r}{d\bar{z}})^2 = (1-\psi_r^2-\bar{v}^2)^2 
\label{eqnpsir}
\end{equation}
where $\bar z = \Lambda z$. 

Now, from Eq. (\ref{gpeg}), the normalized condensate density $\bar \rho_{g} = \rho_{g}/\rho_{g}^0$  for the GPE is
\begin{equation}
\bar{\rho}_g = \psi_r^2+ \bar{v}^2
\label{rgbar}
\end{equation}
Differentiating Eq. (\ref{rgbar}) with respect to $\bar z$, and using
Eq. (\ref{eqnpsir}), it is easy to show that $\bar{\rho}_g$ satisfies the 
following equation.

\begin{equation}
(\frac{d \bar{\rho}_g}{d\bar{z}})^2 = 
( 1-\bar{\rho}_g)^2 ( \bar{\rho}_g-\bar{v}^2 ).
\label{eqnrg}
\end{equation}

The above equation describes the spatial evolution 
of the condensate density whose order parameter
 obeys the GPE, in a frame moving with velocity $v$.
This equation  is identical in form to 
Eq. (\ref{eqnrs}) for the half-filled HGPE, and hence 
the condensate density $\rho_g$ for the GPE (which is equal to to 
the particle density) has the same 
functional form as the condensate density $\rho_s$ for the HGPE. 
We would like to emphasize that this mapping between the GPE
and the HGPE is valid {\it provided one considers only 
the condensate fraction} of the HCB. Since the 
half-filling case for particle density in
HGPE corresponds to quarter-filling for the condensate density, 
the corresponding GPE density for this mapping
to be applicable is $\rho_g^0=1/4$.

The mapping between the GPE and the HGPE discussed 
above provides the relationship that the width of the
soliton for the condensate density  
 in the GPE  and the HGPE are given, respectively, by
$\Gamma_g = (2\gamma \Lambda)^{-1}$ and 
$\Gamma_{s} = 
(2 \gamma \Lambda/\sqrt{1- 2 \Lambda^2})^{-1}$ \cite{relation}. 

We would like to note that the relationship between the GPE and 
the HGPE applies only to the
condensate density profiles, as the solitons for the corresponding 
condensate wave functions in these two limiting cases 
will differ in their phase jumps.\\ 

\noindent {\bf 5. The BEC  soliton and the 
magnetic soliton in the easy-plane ferromagnet}\\

An interesting consequence of the mapping between the GPE
and the HGPE discussed in the previous section is  
the emergence of a relationship between two distinct  
physical systems, 
the GPE soliton in a BEC and a  magnetic soliton that arises from the 
 {\it quantum}  easy-plane ferromagnet 
(\ref{spinH}). As we discuss below, 
this provides an interesting way to understand
why there are in general two solitary waves in the 
bosonic density of the HCB  system, and why in the special case of half-filling,
the condensate part of the HCB has only one solitary wave 
and {\it why it is dark}.

In magnetism, the  study  of strict solitons and 
solitary waves  in 
various types of quasi-one dimensional {\it classical} spin systems has been 
 a very active field \cite{dejo,kose} for over three decades. 
In the present work,  the 
relationship between the Bose Hubbard model  
  and  a  {\it quantum}  ferromagnetic  
Heisenberg spin Hamiltonian (\ref{spinH}) 
with an {\it exchange anisotropy} and a 
magnetic field will be exploited
to relate the BEC soliton to the magnetic soliton. Note
that the anisotropy is of the easy-plane
type since $g > 0$.

By writing the complex parameter    $\tau$ 
appearing  in the spin coherent representation \cite{Radc}  as 
$\tau = \tan (\theta/2) \exp (i \phi)$, 
with $0 \le \theta\le \pi$ and $0 \le \phi \le 2 \pi$,
the coherent state  average  
$<\hat{S}^+>$ 
in  the  quantum spin-$1/2$  system 
can be calculated in terms of $\theta$ and $\phi$.
By identifying
$\rho_s = |<\hat{S}^+>|^2$
we obtain the following relation between the condensate density 
$\rho_s$ and the {\it classical} spin variables: $S_x,S_y,S_z$,
where  $\theta$ and $\phi$ appear as the polar and 
azimuthal angles of a classical spin vector ${\bf S}$.

\begin{equation}
\rho_s  =  (S_x^2+S_y^2)/4  = [ (1/4)  -  S_{z}^{2}]/4 \label{rhos}
\end{equation}

In addition, the particle density $\rho$, the 
chemical potential $\mu$  and the particle-hole imbalance variable 
$(1 - 2\rho^{0})$  in
the BEC system described by HGPE are related to   
 the spin variables of the magnetic system as follows. 

\begin{eqnarray}
\rho & = & 1/2-S_z \label{rho} \\
\mu & = & 2 g \rho^{0} =  g(1-2S_z^{0})\\
\,\,\,\,\,\,\,\,\,g(1-2\rho^0) & = & (g - \mu) = h_{z},
\label{relation}
\end{eqnarray}

where $h_{z}$ is the 
magnetic field along the z-direction, as seen from the 
XXZ Hamiltonian (\ref{spinH}).
Of the above, the first equation is obtained by taking 
the expectation value of the corresponding hard-core boson operator
to be  spin coherent state average. The last two are obtained
by using their asymptotic relationships in the expression for $\mu$ 
(see below Eq. (\ref{sgpe1})). 

On taking spin-coherent state averages of the
quantum spins  on the lattice and going to the continuum, the evolution
equation for the condensate order parameter  also describes
the evolution of classical spins, where the asymptotic
 density  controls  the external
transverse field in the spin system. The solutions
discussed here  correspond
to the half-filled case where this external
field is tuned to zero.

As a consequence of the relationship
(\ref{rho}),  a soliton solution of
the total particle density $\rho $  
corresponds to a nonlinear excitation
 of the $z$-component of the spin.
On the other hand, the soliton of the condensate density describes a
 nonlinear excitation of the in-plane spin, and relates to
the {\it square of the in-plane magnetization},  due to the
relationship (\ref{rhos}).

As discussed in our earlier paper\cite{prl}, the existence of the 
two solitons in HCB has its root in the fact 
that both particles and holes play
 equal roles in deciding the dynamics of the system. It is noteworthy
that  the XXZ  Hamiltonian (\ref{spinH}) provides an alternative 
means to understand
the existence of the two solitons. 
The spin Hamiltonian (\ref{spinH}) contains terms with two 
competing effects, the easy plane anisotropy $g$ that tends
to align spins in the $xy$-plane and the the transverse 
field $h_{z}$ that favors a spin alignment along the $z$-axis.
The ground state of the system consists of spins 
aligned on a cone that makes an angle $\theta_0$ with the
$z$-axis where $\theta_0$ is determined by the 
background density of the atomic cloud given by 
$\rho_0=1/2 - S_{z}^{0} =  1/2- (1/2) \cos \,\,\theta_0$.

Interestingly, in the half-filled case $\rho ^{0} = 1/2$, 
the particle-hole imbalance $\delta_{0} = (1 - 2 \rho^{0})$ is 
zero, and $h_z$ vanishes.
In this case, the ground state consists of spins in the $xy$-plane. 
Therefore, the solitary wave excitation resulting
in a localized change in the density makes the spins move 
 out of the plane. 
In view of the fact that there
is no clear preferred direction, disturbances 
above and below the easy plane are equally preferred. 
In the corresponding particle density in HCB, the
dark and  antidark solitons are therefore
mirror images of each other, as seen from Eq. (\ref{fsol}).

In contrast, as seen in Eq. (\ref{rhos}), 
the condensate density of HCB is related to
the {\it square of the in plane magnetization}, 
(which maps to the GPE density profile
 with $\rho_{g}^{o} = 1/4$) resulting in 
a {\it single} soliton both in the  GPE and in the condensate fraction
of the  HCB. Such a soliton has to be a dark soliton, since the 
out of plane component of the
spin {\it decreases}  the total magnetization of the system.

In summary, soliton characteristics  in the BEC 
of bosons in  weakly and strongly interacting regimes are 
 expected to be quite distinct. In cold atom 
laboratories where the interaction between
 atoms can be varied, one expects 
significant changes in the dynamics as one tunes
the scattering lengths of the bosonic atoms. 
Therefore, our result illustrating the similarities 
between the solitons  in
these two different regimes in the special half-filled 
case is important. Furthermore, by exploiting 
the relationship between the  HCB and spin systems,
the connection between the  GPE and HGPE  solitons paves the way 
for relating them to magnetic solitons  in a
 ferromagnetic spin chain with easy-plane exchange anisotropy.
Thus the HGPE soliton at half-filling and the 
above magnetic soliton can be regarded as two other incarnations 
of the GPE soliton.
These studies open a new way to understand and interpret
many  important characteristics of solitons 
in these systems, including the fact that it provides
an intuitive understanding of why the GPE soliton is {\it dark}.

It is to be noted that  it is the density in BEC that has 
a solitonic behavior. 
Thus by designing 
a quasi-one-dimensional optical lattice which incorporates 
the hard-core boson constraint of no double occupancy,
and by loading them with bosonic atoms appropriately so as to 
simulate a  half-filled lattice, 
it would be of interest to
study  soliton propagation of density in this strongly repulsive system, 
and investigate its relationship to the GPE dark soliton 
which  is the hallmark of a weakly repulsive system.\\

\noindent {\bf Acknowledgments}\\

Research of IIS is
supported by the grant N00014-09-1-1025A by the Office of
Naval Research, and the grant 70NANB7H6138, Am 001 by the National Institute
of Standards and Technology.
RB thanks the Department of Science and Technology, India for financial support.


\begin{thebibliography}{99}

\bibitem{daux} T. Dauxois and M. Peyard,  Physics of Solitons, Cambridge University Press, Cambridge,  2006.

\bibitem{ablo} M. J. Ablowitz and H. Segur, Solitons and Inverse Scattering Transform,  SIAM, Philadelphia, 1981.

\bibitem{book} L. Pitaevskii and S. Stringari, 
Bose-Einstein Condensation,  
Oxford University Press, Oxford,  2003.

\bibitem{tsuz} T. Tsuzuki, J. Low. Temp. Phys.  4 (1971)  441.

\bibitem{burg} S. Burger {\em et al.}, Phys. Rev. Lett. 
 83 (1999)  5198 ; 
J. Denschlag {\em et al.}, Science   287 (2000)  97. 

\bibitem{kol} E. Kolomeisky, T. J. Newman, J. Straley,
and X. Qi,  Phys. Rev. Lett.  85 (2000) 1146 .

\bibitem{lew}K. V. Krutitsky, J. Larson and M. Lewenstein, 
arXiv: 0907.0625.

\bibitem{Carr} R. V. Mishmash, I. Danshita, C. W. Clark, and L.D. Carr, 
Phys. Rev. A 80 (2009) 053612.

\bibitem{lang} J. S. Langer,  Phys. Rev.   167 (1968)  183 .


\bibitem{radha1} R. Balakrishnan, R. Sridhar, and R. Vasudevan, 
Phys. Rev. B  39 (1989) 174 ; R. Balakrishnan, 
Phys. Rev. B  42 (1990)  6153 . 

\bibitem{mats} T.  Matsubara and H. Matsuda, 
Prog. Theor. Phys.  16 (1956)   569 ;
S.Sachdev , Quantum Phase Transitions, 
Cambridge University Press, Cambridge, 1999.

\bibitem{prl} R. Balakrishnan, I. I Satija and C. W. Clark, Phys. Rev. 
Lett.  103 (2009) 230403 .

\bibitem{anti} For the definition of an antidark soliton, see  
Y. S. Kivshar and V. V. Afanasjev, 
Phys. Rev. A  44  (1991)  R1446 .

\bibitem{Baiz} B. B. Baizakov {\em et al}. J. Phys. B: 
At. Mol. Opt. Phys. 42 (2009) 175302;
G. Gligoric {\em et al}, Phys Rev A 78 (2008)  063615.

\bibitem{Radc} J. M. Radcliffe, J. Phys. A  4  (1971)  313 .

\bibitem{relation} In this relationship, it should 
be noted that the dimensionless
parameter  $\Lambda$  is $c_g/c_{0}$ for the GPE and $c_s/c_0$ for 
the HGPE.  Likewise,
 in the parameter 
$\gamma = (1 -\bar v^2)^{1/2}$, the dimensionless soliton speed 
$\bar v$ is $v/c_g$ for the GPE and $v/c_s$ for the HGPE.   

\bibitem{dejo} L. J. de Jongh and A. R. Miedema, 
Adv. Phys.  23 (1974)  1 .

\bibitem{kose} A. M. Kosevich, B. A. Ivanov and
A. S. Kovalev, Phys. Rep. 194 (1990)  117.

\end{thebibliography}
\end{document}